\documentstyle[prl,aps,epsfig]{revtex}

\begin{document}
\title{\bf
      Controlling spatiotemporal chaos via small external forces\\
      } \vskip 5mm 
\author{\small Shunguang Wu$^{a,b}$, Kaifen He$^{a,b}$ and Zuqia Huang$^{b}$\\
\small $^a$CCAST ( World Laboratory ), P. O. Box 8730, 100080, Beijing, China\\
\small $^b$Institute of Low Energy Nuclear Physics,Beijing Normal University, 
          100875, Beijing, China} 
\maketitle
\begin{quote}
\parbox{16cm}{\small
The spatiotemporal chaos in the system described by
a one-dimensional nonlinear drift-wave
equation is controlled by directly adding a periodic force
with appropriately chosen frequencies.
By dividing the solution of the system into a carrier steady
wave and its perturbation, we find that
the controlling mechanism can be explained by a slaving principle.
The critical controlling time for a perturbation mode
increases exponentially with its wave number.\\
\quad\\
PACS numbers:05.45+b, 52.35.Kt, 52.35.-g
}
\end{quote}

\section{Introduction}
Spatiotemporal  chaos(STC) can  appear in a large variety of systems such as
hydrodynamic systems, plasma devices, laser systems, chemical reactions,
Josephson junction arrays and biological systems.
Because of the potential applications, controlling STC in these systems
has been given much attention by scientists and technologists in recently
years\cite{Hu1}-\cite{Mont}.
Generally speaking, there are two kinds of methods to
control STC: the feedback control\cite{Hu1}-\cite{Parm}
and the nonfeedback one\cite{Aran}-\cite{Mont}.
For the former one, it needs a fast responding feedback system that produces
an external signal in response to the system's dynamics.
On the contrary, for the later one, the form and the amplitude
of the external control signal can be adjusted numerically or experimentally 
by directly observing the 
response of the system to the applied signal.
The advantages and disadvantages of the two kinds of methods are:
for the feedback one, 
the reference state  is always an unstable trajectory locating in the chaotic 
attractors, 
the controlling input is very small if the
system can be well controlled, 
in this way, however, one must know some prior knowledge of the
system before controlling;  In contrast,
for the nonfeedback one, 
it does not need any prior knowledge of the system,
it is very easy in practice, and particularly convenient for experimentalist,
but the target states are not the unstable
periodic orbits of the chaotic attractors, hence, the controlling input
does not vanish when the system is under control.

In this letter, we will give an example of controlling STC by adding a
small external force on a system and discuss the mechanism.
In fact, the methods of adding a pre-determined driving force or modifying
an accessible parameter in a pre-determined way on the system to control
chaos might be firstly used by Alekseev and Loskutov\cite{Alek},
who studied how to
control a chaotic model of a water ecosystem by small periodic perturbations.
After Alekseev and Loskutov, Lima and
Pettini\cite{Lima} successfully used the parametric perturbation method to
eliminate chaos in Duffing-Holmes oscillator. Some experimental works about
the method have been done by Azevedo\cite{Azev},Fronzoni\cite{Fron},
Ding\cite{Ding} and Ciofini\cite{Ciof}.
Moreover,
Braiman and Goldhirsch\cite{Brai} directly added a weak external periodic
force to Josephson junction system to suppress chaos even in nonresonant
circumstance. Recently, the method has been intensively and extensively used
to control chaos not only for nonautonomous systems but also for autonomous
systems and iterated maps\cite{Meth}.
However, within our knowledge, the method has not ever been used
to control STC.
This letter will give the first example to use the method
to control STC in a partial differential equation systems, and find for the
first time that the mechanism can be explained by the slaving principle.

The letter is organized as follows. First,
we will briefly introduce the driven-damped nonlinear drift-wave
equation which we used as the model equation.
Then, the control method and the numerical simulation results
will be given. Next, we will analyze the controlling mechanism.
Finally, a discussion and conclusion will be given.

\section{Model}
The model we used in this letter is a one-dimensional nonlinear drift-wave
equation driven by a sinusoidal wave. It reads\cite{He1}
\begin{equation}
\label{s21}
   \frac{\partial \phi}{\partial t}+a\frac{\partial^3 \phi}{\partial t \partial x^2}
   +c\frac{\partial \phi}{\partial x}+f\phi\frac{\partial \phi}{\partial x}
   =-\gamma\phi-\epsilon \sin(x-\Omega t),
\end{equation}
where the {\rm $2\pi$}-periodic boundary condition,$\phi(x+2\pi,t)=\phi(x,t)$,
is applied. 
In this study, we fix the parameters $a=-0.2871$, $\gamma=0.1$, $c=1.0$
and $f=-6.0$
, only the forcing strength $\epsilon$ and phase speed $\Omega$ of the
sinucoidal wave are
the control parameters. 
Without the dissipation and  driving terms($\gamma=0$, $\epsilon=0$),
Eq. (\ref{s21})
describes the nonlinear drift-wave in magnetized plasmas \cite{Ora}.
After introducing the driving and dissipation terms, the competition among
dispersion, dissipation, driving and nonlinearity  makes the system display
rich dynamic phenomena\cite{He1}\cite{He2}. 

In plasma physics, an integral quantity $E(t)$, which
is well known as the "energy" of the system,
\begin{equation}
\label{s22}
   E(t)=\frac{1}{2\pi}\int_{0}^{2\pi} \frac{1}{2} [ \phi^2(x,t)
       -a(\frac{\partial \phi}{\partial z})^2 ] dz,
\end{equation}
is conveniently used to monitor the dynamics of the system.  $E(t)$ is 
a positive constant at $\epsilon=\gamma=0$ and monotonously decreases to
zero by the dissipation if $\epsilon=0$ and $\gamma>0$.

In the present paper, we focus on STC states.
On the parameter plane $\epsilon-\Omega$, the dynamic behavior of the
system has been extensively studied by one of us(He)\cite{He1,He2}.
According to Fig. 6(b) in Ref. \cite{He2}, one can find
that in certain regions of $\Omega$, when the driving $\epsilon$
is strong enough the system displays a STC state.
It is necessary to point out that so called space-time chaos
is an ambiguous terminology.  There is evidence to show that the spatial
incoherence may be caused by different physical effects.
In previous work\cite{Hu2}, Hu and He reported an example of controlling
one type
of STC states by feedback method, its erratic spatial behavior is due to
the overlapping of the regimes  of the Hopf instability of different modes.
In the present work, we will focus our attention
on a STC state, e.g. at  $\epsilon=0.22, \Omega=0.65$, the transition
to STC is caused by a crisis\cite{He2}.

\section{Methods and numerical simulation results}
For controlling STC in the system, we directly add a small temporally
periodic signal on
the right hand side of Eq. (\ref{s21}). Thus Eq. (\ref{s21}) is changed to
\begin{eqnarray}
\label{s31}
   \frac{\partial \phi}{\partial t}+a\frac{\partial^3 \phi}{\partial t \partial x^2}
   +c\frac{\partial \phi}{\partial x}+f\phi\frac{\partial \phi}{\partial x}
   =-\gamma\phi-\epsilon \sin(x-\Omega t) 
    +\eta \cos(\omega t),
\end{eqnarray}
where $\eta$ and $\omega$ are the control parameters.  

The pseudospectral method\cite{Orsz} is used to simulate Eq. (\ref{s31}).
In the numerical simulation, we divide the space into 128 points.
256 and 512 points have also been used,
the results do not show qualitative difference. Therefore, the 128 points
are used in our work. 

In order to find suitable $\eta$ and $\omega$ to control STC,
we first calculate the power spectrum of $\phi(x_0,t)$,
denoted by $S(\omega)$, at a  fixed point $x=x_0$ in space.
From the power spectrum $S(\omega)$ one can find 
the characteristic frequency region of the chaotic attractor.
For example, Fig. 1(a)
shows the spectrum of the STC state to be controlled,
from this figure one can see
that except for the frequency of driving wave($\Omega=0.65$), there are many
peaks in the region $\omega \in [0.2,0.9]$.
Our experience shows that the appropriate $\omega$ for a
successful control should be chosen in this region. Now, fix an
$\omega$ in this region, one can try to find a suitable $\eta$.
We find some  windows in $\omega - \eta$ plane,
 with which parameters the turbulent motion of the STC state
 can be successfully suppressed. 
 The results are given in
Fig. 2, in which 
the circles stand for the suitable parameter points for controlling STC 
in system (1), while  the crosses  indicate the parameter points at
which the turbulence are failed to be controlled.
In the following,
we choose $\omega=0.756$ and $\eta=0.1$ as an example.
To compare the
dynamic behavior of the system before and after the control, the power spectrum,
energy and spatiotemporal patterns are calculated.
Figure 1(b) shows the power spectrum $S(\omega)$ at a
fixed space point($x=\pi$)
after the control. In contrast  to the wide noisy spectrum
in Fig. 1(a), here only a few lines are left.
In Fig. 3,
We give the time series of the energy $E(t)$, one can see it is chaotic before 
control,
which becomes harmonic oscillations after a while when the control signal is added.
The spatiotemporal patterns of the system before and after control
are shown in Fig. 4(a) and Fig. 4(b), respectively, obviously the turbulence
is strongly suppressed .

\section{Mechanism and  Scaling behavior}
To analyze the mechanism of the control method introduced above,
firstly, let us observe the solution of Eq. (\ref{s31}) in the frame  of
the driving wave. By setting $z=x-\Omega t$ and $\tau=t$ , Eq. (\ref{s31})
can be rewritten as
\begin{equation}
\label{s40}
   \frac{\partial}{\partial \tau}(1+a\frac{\partial^2}{\partial z^2})\phi
   =\Omega \frac{\partial}{\partial z}(1+a\frac{\partial^2}{\partial z^2})\phi
   -c\frac{\partial \phi}{\partial z}
   -f\phi\frac{\partial \phi}{\partial z}
   -\gamma\phi-\epsilon \sin z + \eta \cos(\omega \tau).
\end{equation}
Then, we divide a solution of Eq. (\ref{s31}) into a carrier steady
wave $\phi_0(z)$, and it's perturbation $\delta \phi(z,t)$, i.e.,
\begin{equation}
\label{s41}
    \phi(z,t)=\phi_0(z)+\delta \phi(z,t).
\end{equation}
Here $\phi_0$ satisfies the steady equation $\partial \phi_0/\partial \tau=0$,
i.e.,
\begin{equation}
\label{s42}
   \Omega \frac{\partial}{\partial z}(1+a\frac{\partial^2}{\partial z^2})\phi_0
   -c\frac{\partial \phi_0}{\partial z}-f\phi_0\frac{\partial \phi_0}
   {\partial z}   -\gamma\phi_0-\epsilon \sin z=0.
\end{equation}
Insert Eq. (\ref{s41}) into Eq. (\ref{s40}), and using Eq. (\ref{s42}), one has
\begin{eqnarray}
\label{s43}
  \frac{\partial}{\partial \tau}(1+a\frac{\partial^2}{\partial z^2}) \delta \phi
   &=&\Omega \frac{\partial}{\partial z}(1+a\frac{\partial^2}{\partial z^2})\delta \phi 
   -c\frac{\partial \delta\phi}{\partial z}
   -f\phi_0\frac{\partial \delta\phi}{\partial z}
   -f\delta\phi\frac{\partial \phi_0}{\partial z}
   -f\delta\phi\frac{\partial \delta\phi}{\partial z}  \nonumber\\
   &&-\gamma\delta\phi
   +\eta \cos(\omega \tau).
\end{eqnarray}
By making the Fourier transformation of $\phi_0(z)$ and $\delta\phi(z,t)$ as
\begin{equation}
\label{s44}
   \left\{
       \begin{array}{l}
         \phi_0(z)=A_0+lim_{N\rightarrow \infty}\sum_{k=1}^{N} A_k\cos(kz+\theta_k), \\
         \delta \phi(z,t)=b_0(t)+lim_{N\rightarrow \infty}\sum_{k=1}^{N} b_k(t)\cos(kz+\alpha_k(t)),
       \end{array}
   \right.
\end{equation}
Eq. (\ref{s42}) is transformed to a set of infinite algebraic equations, and
Eq. (\ref{s43})  a set of infinite ordinary differential
equations
\begin{equation}
\label{s45}
  \left\{
  \begin{array}{l}
    \frac{d b_0}{dt}    = -\gamma b_0+ \eta \cos(\omega t),\\
    \frac{d b_k}{dt}    = N_k(t)+ f k b_0 A_k \sin(\theta_k-\alpha_k)/(1-ak^2),\\
    \frac{d \alpha_k}{dt}= M_k(t)-\frac{f k (A_0+b_0)}{(1-ak^2)}
    -\frac{f k b_0 A_k\cos(\theta_k-\alpha_k)}{(1-ak^2)b_k},
  \end{array}
  \right.
\end{equation}
where
\begin{eqnarray}
N_k(t)=&-&\frac{\gamma b_k}{1-ak^2}+\frac{fk}{2(1-ak^2)} \nonumber\\
       &\times& \{ \sum_{l+l'=k}[A_lb_{l'}\sin(\theta_l+\alpha_{l'}-\alpha_k)
                         +\frac{1}{2}b_lb_{l'}\sin(\alpha_l+\alpha_{l'}-\alpha_k)] \nonumber\\
       &+&  \sum_{l-l'=k}[A_lb_{l'}\sin(\theta_l-\alpha_{l'}-\alpha_k)
                         +\frac{1}{2}b_lb_{l'}\sin(\alpha_l-\alpha_{l'}-\alpha_k)] \nonumber\\
       &+&  \sum_{l'-l=k}[A_lb_{l'}\sin(-\theta_l+\alpha_{l'}-\alpha_k)
                         +\frac{1}{2}b_lb_{l'}\sin(-\alpha_l+\alpha_{l'}-\alpha_k)] \}, \nonumber\\
M_k(t)=&-&k ( \frac{c}{1-ak^2}-\Omega ) - \frac{kf}{2(1-ak^2)b_k} \nonumber\\
       &\times& \{ \sum_{l+l'=k}[A_lb_{l'}\cos(\theta_l+\alpha_{l'}-\alpha_k)
                         +\frac{1}{2}b_lb_{l'}\cos(\alpha_l+\alpha_{l'}-\alpha_k)] \nonumber\\
       &+&  \sum_{l-l'=k}[A_lb_{l'}\cos(\theta_l-\alpha_{l'}-\alpha_k)
                         +\frac{1}{2}b_lb_{l'}\cos(\alpha_l-\alpha_{l'}-\alpha_k)] \nonumber\\
       &+&  \sum_{l'-l=k}[A_lb_{l'}\cos(-\theta_l+\alpha_{l'}-\alpha_k)
                         +\frac{1}{2}b_lb_{l'}\cos(-\alpha_l+\alpha_{l'}-\alpha_k)] \}, (k=1,2,\dots,N). \nonumber
\end{eqnarray}

In numerical simulations, the mode equations for ${A_k, \theta_k}$ and
${b_k, \alpha_k}$ can be solved simultaneously with truncation.
The proper $N$ should satisfy the condition that
the modes $k>N$ give negligible contributions to the actual solution of the
equation\cite{Hu2}.  For the present parameters we use
$N=13$, i.e., the 27-dimensional ordinary differential
equations will be solved. With these mode equations one can obtain
qualitatively the same results as the direct simulating Eq. (\ref{s31}).

Now let us analyze the numerical results of the solution of Eqs. (\ref{s45}).
From the time evolution curves of $b_k$, one can see
that, for every mode $k$, there exists a turning point $\tau_k^{*}$,
before which $b_k$ shows chaotic motion with larger amplitude,
while after which it becomes a quasi-periodic vibration
with much smaller amplitude.
As an example, Fig. 5(a) depicts the time evolution of $b_3$  and $b_{11}$
in the beginning 50 time units after the control signal is added and
another 50 time units when they behave quasi-periodically.
In order to determine $\tau_k^{*}$, let us 
calculate the relative average amplitude value of each mode
\begin{equation}
r_k(t)=(b_k^{*}(t)-\bar {b_k})/ \bar {b_k},
\end{equation}
where $\bar b_k$ is the time average value of $b_k$ in the
quasi-periodic states,
$b_k^{*}$ is the points only on the Poincar{\rm \'e} section\cite{Not1}.
As an example, the results of $r_8(t)$,$r_{10}(t)$ and $r_{12}(t)$
are shown in Fig. 5(b),  one can see that, after adding the periodic control
signal, $r_k(t)$ starts to decrease
exponentially,  but they never tend to zero. After
a certain time point, the oscillation level of each $r_k$ no longer
decrease.

Next, we will study the dependence of the critical
point $\tau_k^{*}$ on the
mode number $k$. In the decreasing phase of $r_k$, its maxima  
decreasing linearly in the logarithmic diagram, while in the quasiperiodic phase,
the maxima of $b_k$  are about a constant. The intersection point
of the two straight lines gives $\tau_k^{*}$. In Fig. 5(c), we plot
$\tau_k^{*}$ as a function of mode
number $k$ in the semilogarithmic plane, the points can be well
fitted by a
straight line. Consequently we obtain a scaling law
\begin{equation}
\label{s46}
\tau_k^* \propto e^{\beta k},
\end{equation}
here by the least square method we get $\beta=0.1169$.  When amplitude
$b_k$'s transit to the quasi-periodic phase, the mode phase
,$\alpha_k$, of the originally chaotic mode also become very quiescent. 
We can also determine the turning points $\tau_k^*$ from the evolution of 
$\alpha_k$, the same scaling law as Eq. (\ref{s46}) is obtained,
with $\beta=0.1158$.

From these results we conclude that, by using our time-periodic signal 
to control a STC state, the modes$\{b_k, \alpha_k\}$ will be controlled 
one by one from the smallest wave number $k$ to the highest one. The 
critical time for a mode to be controlled increases exponentially with 
the wave number $k$. This phenomenon can be
understood by the slaving principle in synergetics\cite{Haken}.
From the first equation of Eqs. (\ref{s45}), one can see that,
in fact, the external force is
added on the mode $k=0$, the evolution of $b_0$ is uniquely determined
by the applied control signal and the dissipation rate $\gamma$.
Its motion does not influenced by the other $k \neq 0$ modes.  
The motion of $b_0$ is adjusted to periodic one very quickly after 
the control signal is added. Then the other modes
$k \neq 0$ are slaved by $b_0$. This fact can be seen from Eqs.(\ref{s45})
for $k\neq 0$. In both equations for $b_k$ and $\alpha_k$, there is a term
depending on $b_0$, which is the contribution of the mode-mode couplings
between $b_0$,$\{A_k, \theta_k\}$ and $\{b_k,\alpha_k\}$. Through these terms
$b_0$ plays the slaving role.
Meanwhile, since the modes of the steady
wave $\phi_0(z)$, $\{A_k, \theta_k\}$ are constants in the frame
of $\{z,\tau\}$. the modes $\{b_k, \alpha_k\}$ also
play a significant role in the solving process.
Moreover, the numerical results tell us that $b_0$ is a
slowly relaxing variable and the other modes are fast ones
in the beginning stage  when the control signal is just added.
Thus, $b_0$ plays an important role as an order parameter in the system,
and it is affected by the control force. When the control signal is added,
$b_1$ and $\alpha_1$ are immediately slaved by $b_0$.
Since the effect of $b_0$ on $\{b_k, \alpha_k \}$ is through
$\{A_k, \theta_k\}$ and $A_{k}$ decrease with $k \rightarrow \infty$
, the slaving intensity is not unform for
different modes. From Eqs. (\ref{s45}), one can see
that the smaller the
mode $k$, the higher the slaving intensity it has. 
So the mode $k=1$ can be controlled firstly. Then, it starts to slave the
modes $k=2,3,...$, until all the modes are tamed. In the whole
process, the system changes from an extremely disordered(a STC state) into
an ordered one by the self-organization principle.

\section{Conclusion and discussion }
In conclusion, we have successfully suppressed very turbulent
 STC  states by directly adding
a temporally periodic force on the system. 
The appropriate input frequency can be chosen by analyzing the spectrum
of the turbulence.
In contrast to the feedback method for which
one needs the knowledge of the target state,
the present method is easy to be realized in practical
experiments. The controlling mechanism is studied  by investigating
the evolution of perturbation wave to the carrier steady wave.
We find that the perturbation modes are controlled
one-by-one from long to short wave length by a slaving principle,
and the critical controlling time
for the mode increases exponentially with its wave number $k$. Since the
slaving  effect is very important in nonlinear science, these results
may be interested in by scientists and engineers. The schemes of how to select
the controlling frequency may also be helpful for the experimentalists.

\begin{center}
{\large {\bf ACKNOWLEDGMENTS}}\\
\end{center}
The first author(Wu) would like to thank Drs. Baisong Xie and Luqun Zhou
for useful discussions.
This work is supported by the National Natural Science Foundation
of China
under grant No. 19675006, No. 19835020 and the Research Funds for the
Doctoral Program of Higher Education under the grant No. 98002713.

\newpage
\begin{figure}
\centerline{\psfig{figure=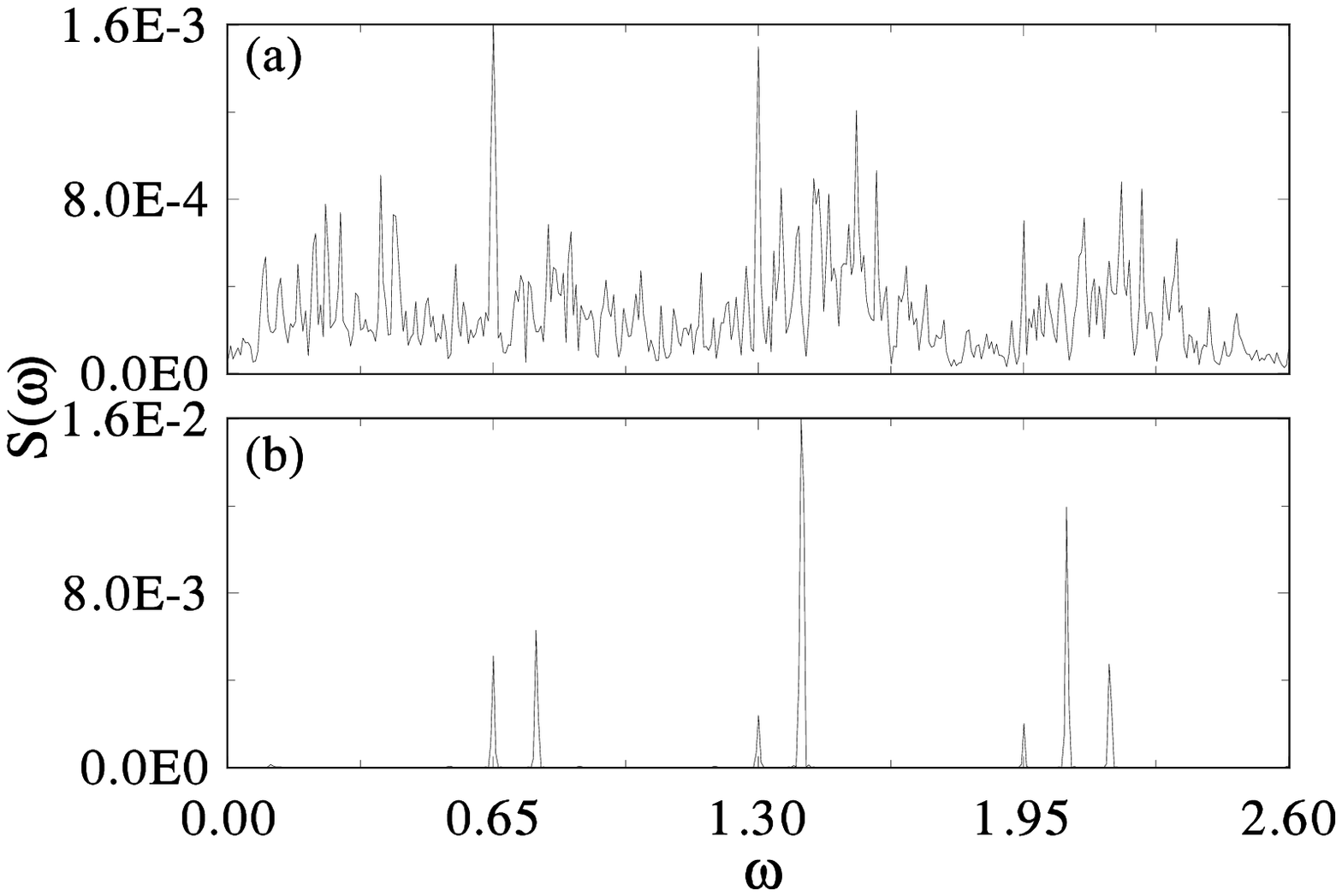,width=8cm,angle=0}}
\caption{
Power spectrum of potential at fixed $z$($z=\pi$)  before(a) and 
after(b) control.
The time interval,$\Delta t$, between consecutive samples is $0.5$, The
1024 points are used to do the FFT, the total sample points are $4096$.
}
\end{figure}

\begin{figure}
\vspace{-1cm}
\centerline{\psfig{figure=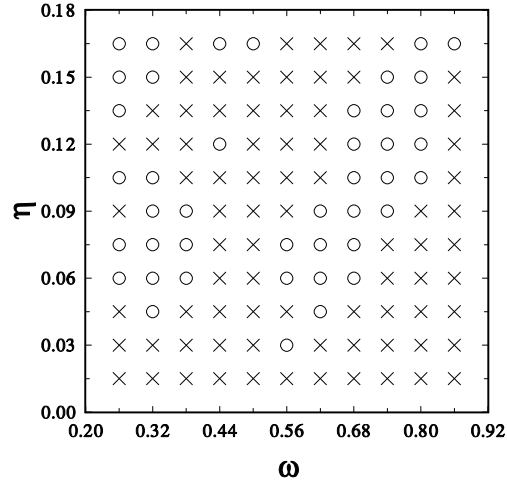,width=10cm,angle=0}}
\caption{
Global view of control parameters on the plane $\omega-\eta$.
Points $\circ$ and $\times$ stand for the suitable and unsuitable
parameter points for controlling STC in system (1),respectively.
 }
\end{figure}

\begin{figure}
\centerline{\psfig{figure=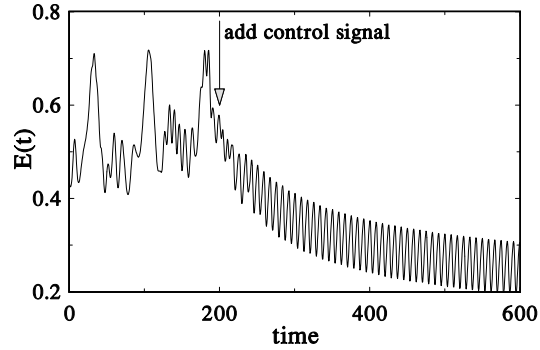,width=8cm,angle=0}}
\caption{
The energy E(t) as a function of time before and after control.\hspace{3cm}
 }
\end{figure}

\begin{figure}
\hspace{1cm}
\centerline{\psfig{figure=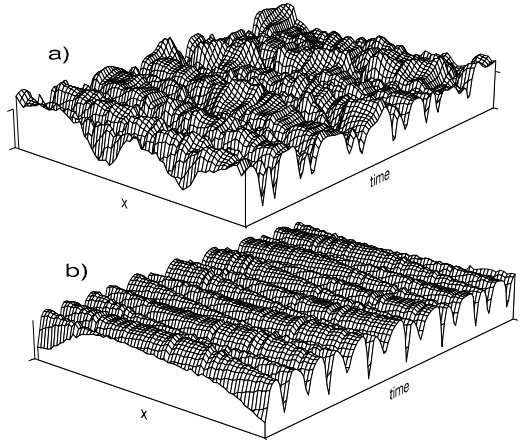,width=12cm,angle=0}}
\vspace{-3cm}
\caption{
Spatiotemporal patterns of $\phi(z,t)$ before(a) and after(b) control.\hspace{3cm}
 }
\end{figure}

\begin{figure}
\centerline{\psfig{figure=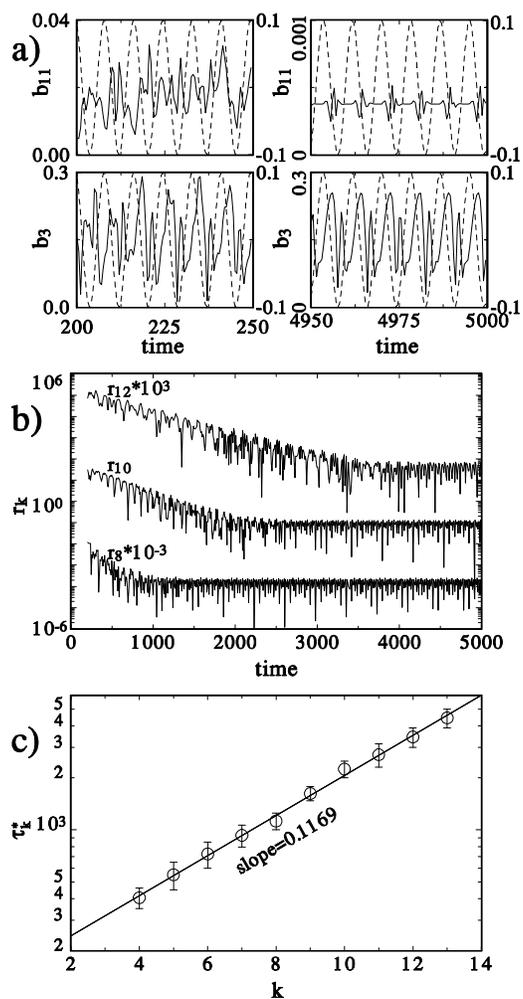,width=20cm,angle=0}}
\caption{
(a) The evolution of the amplitudes, $b_k(t)(k=3,11)$, for the
perturbation wave, the dashed lines scaled by the right axises indicate
the external force. (b) The evolution of $r_k(k=8,10,12)$.
(c) The scaling law of $\tau_k^*$.
}
\end{figure}

\begin{thebibliography}{99}
\bibitem{Not1} Here, the Poincar{\rm \'e} section is
               the time at which  the control signal has a minimum value.
\end{thebibliography}
\end{document}